\documentclass[a4paper]{jpconf}
\usepackage{amsmath}
\usepackage{amsfonts}
\usepackage{amssymb}
\usepackage{anyfontsize}
\usepackage{srcltx}
\def\rf#1{(\ref{eq:#1})}
\def\lab#1{\label{eq:#1}}
\begin{document}
\title{Symmetries and hamiltonians of Ince's  XXXVIII and XLIX equations
}
\author{
V.C.C. Alves$^{1}$,
 H. Aratyn$^{2}$, 
J.F. Gomes$^{1}$
 and A.H. Zimerman$^{1}$
}
\address{
$^{1}$
Instituto de F\'{\i}sica Te\'{o}rica-UNESP\\
Rua Dr Bento Teobaldo Ferraz 271, Bloco II,\\
01140-070 S\~{a}o Paulo, Brazil}
\address{$^2$ Department of Physics \\
University of Illinois at Chicago\\
845 W. Taylor St.\\
Chicago, Illinois 60607-7059
}
\ead{victor.cesar@unesp.com, aratyn@ift.unesp.br, francisco.gomes@unesp.br, zimerman@ift.unesp.br}
\begin{abstract}
We discuss symmetries of Hamiltonians 
of I$_{38}$ and I$_{49}$  equations that appear
on Ince's list of fifty second-order differential equations with 
Painlev\'e property. 
This study is informed by structure of Weyl symmetries 
of Painlev\'e P$_{III}$ and  mixed Painlev\'e
P$_{III-V}$ equations and provides insights into differences between
the symmetries of Painlev\'e equations and symmetries of solvable equations
on Ince's list.
\end{abstract}

\section{Introduction}

Symmetry group analysis has been of crucial importance for
studies of Painlev\'e equations and
the singular behavior of solutions of second-order differential equations
on the complex plane. In addition to the
celebrated $6$ Painlev\'e equations, there are also other $44$ 
ordinary second-order differential equations with solutions that have
no movable critical point other than poles. 
These equations presented in Ince's book \cite{Ince} are solvable,
meaning that their solutions are expressible in terms
of known functions. Comparing with literature on Painlev\'e equations,  the
Hamiltonian structure and symmetries of solvable equations on Ince's
list attracted much less attention with notable exceptions of 
few recent publications \cite{Levi,Noshchenko,Sasano-Ince}.
The present study fills this gap by studying
symmetries of  equations  I$_{38}$ and I$_{49}$ :
\begin{align}
I_{38} &:\, y_{xx} = \left( \frac{1}{2y}+\frac{1}{y-1} \right) y_x^2  
+ y(y-1)
 \left( {\cal A}(y-1) 
+{\cal B}  \frac{y-1}{y^2}+\frac{{\cal C} }{y-1}
+  \frac{{\cal D}}{(y-1)^2} \right)
\lab{Ince:38} \\
I_{49} &:\,y_{xx}=\left(
\frac{1}{y}+\frac{1}{y-1}+\frac{1}{y-\mathcal{A}}\right)  \frac{y_x^2}{2}
+ y (y-1)(y-\mathcal{A})
\left( \mathcal{B}+ \frac{\mathcal{C}}{y^2}+\frac{\mathcal{D}}{(y-1)^2}
+\frac{\mathcal{E}}{(y-\mathcal{A})^2} \right)
\lab{Ince:49}
\end{align}
from Ince's list \cite{Ince,Levi,Noshchenko,Sasano-Ince}.
For the full understanding of their symmetries it is instructive to
first study how their structures emerge in the context of 
 P$_{III-V}$ model \cite{AGZ2016,AAGZ2018}. 
Note that Ince's equation I$_{38}$  \rf{Ince:38} with $\mathcal{D}=0$ 
is equivalent to Ince's  equation I$_{49}$   \rf{Ince:49} with 
$\mathcal{A}=1$.
\newpage
These two incomplete forms of  \rf{Ince:38} (with
$\mathcal{D}=0$) or  \rf{Ince:49} (with
$\mathcal{A}=1$) equations emerge for 
two special values of the parameters : 
\begin{equation} 
\text{1.} \;\; r_1=0, J=1 \qquad 
\text{2.} \;\;r_0=0, J=-1
\lab{special-limits}
\end{equation}
of the P$_{III-V}$ model: 
\begin{equation}
\begin{split}
z q_z &= q \left(q-r_1 \right) \left(2p+r_0 z \right)- 
\left(\alpha_1+\alpha_3 \right) q
+\alpha_1 r_1  +\epsilon_0 r_0 z^{-J}\\
z p_z &=   p \left(p+ r_0 z \right)
\left(r_1 - 2q \right) + 
(\alpha_1+\alpha_3) p - 
\alpha_2  r_0 z -
\epsilon_1 r_1 z^{J+1}\,.
\end{split}
\lab{qzpz}
\end{equation}
defined here in terms of the two first-order Hamiltonian equations. 
These equations depend on a number of parameters
$J, \epsilon_0,\epsilon_1,r_0,r_1$ together with  $ \alpha_j, j=0,1,2,3$ (with
$ \sum_{j=0}^3 \alpha_j =1$) and 
can be obtained from the Hamiltonian:
\begin{equation}
\begin{split}
zH &= q \left(q-r_1 \right) p \left(p+ r_0 z \right)-
\left(\alpha_1+\alpha_3 \right) q p
+\left( \alpha_1 r_1 +\epsilon_0 r_0 z^{-J}\right)p\\
&+\left( \alpha_2  r_0 z+\epsilon_1 r_1 z^{J+1}\right)q\, .
\lab{Hamiltonian}
\end{split}
\end{equation}
The above equations \rf{qzpz} are invariant under an automorphism $\pi$ such that 
\begin{equation}
  \pi(q)=-p/z, \; \pi(p)=(q-r_1)z, \; \pi(\alpha_i)=
 \alpha_{i+1}, \; \pi(J)=-J, \;
 \pi (\epsilon_i)= (-1)^{i}\epsilon_{i+1}, \, i=0,1,
 \lab{pipbarq}
 \end{equation}
together with $\pi (r_i)= r_{i+1}$. The automorphism $\pi$ satisfies $\pi^4=1$.
 
The P$_{III}$  Painlev\'e equation 
\begin{equation} 
P_{III} \;:\; y_{zz}  = - \frac1z y_z+  \frac{y_z^2 }{y}
+{\cal A} \frac{y^2 }{z} +{\cal C} y^3
+ \frac{{\cal B}}{ z}+
\frac{{\cal D}}{y}
\lab{P3}
\end{equation}
emerges from  P$_{III-V}$ model for 
either $r_1 =0$ and $J\ne 1$  or $r_0 = 0$ and $J \ne -1$ and is
invariant under the extended affine Weyl group 
$W [s_0,s_2,\pi_0,\pi_2,\pi^2]$ in the former case and
by $W [s_1,s_3,\pi_1,\pi_3,\pi^2]$ in the latter case. 
The extended affine Weyl group 
$W[ s_0,s_2,\pi_0, \pi_2, \pi^2]$ \cite{AGZ2016} is generated by
transformations 
\begin{align}
\pi_0(q)&= -\frac{\epsilon_0 }{q}, \quad  
\pi_0 (p)= \frac{ 1}{\epsilon_0 } \left( q^2p+\alpha_2 q \right)
\lab{pi0}\\
\pi_0 (\alpha_1+\alpha_3) &=  -2\alpha_2-\alpha_1-\alpha_3, \qquad
\pi_0 (\alpha_2)=  \alpha_2 , \qquad
\pi_0(\alpha_0)= 2-\alpha_0 \nonumber \\
\pi_2(q)&= \frac{\epsilon_0  }{q}, \quad  
\pi_2 (p)= -\frac{ 1}{\epsilon_0 } \left( q^2(p+r_0z)+(1-\alpha_2
-\alpha_1-\alpha_3)q\right)-r_0 z \lab{pi2}\\
\pi_2(\alpha_1+\alpha_3) &=  -2+2\alpha_2+\alpha_1+\alpha_3, \quad
\pi_2(\alpha_2)=  2 - \alpha_2 , \qquad
\pi_2 (\alpha_0)= \alpha_0\nonumber \\
s_2(q)&= q+ \frac{\alpha_2}{p}, \quad  
s_2 (p)= p, \lab{s2def}\\
s_2(\alpha_1+\alpha_3) &=  2\alpha_2+\alpha_1 +\alpha_3, 
\; s_2(\alpha_2)=-  \alpha_2 , \,
\nonumber \\
s_0(q)&= q+ \frac{1-\alpha_2-\alpha_1-\alpha_3}{p+r_0 z}\qquad
s_0 (p)= p\lab{s0def}\\
s_0(\alpha_1+\alpha_3) &=2-2 \alpha_2-\alpha_1-\alpha_3,  \;
s_0(\alpha_2)=  \alpha_2 , \,
\nonumber
\end{align}
that satisfy relations :
\begin{equation} s_i^2=1=\pi_i^2, \; \pi^2 \pi_i \pi^2 =\pi_{i+2}, \;
 \pi^2 s_i \pi^2 =s_{i+2}, \; i=0,2,
\lab{prop1}
\end{equation}
for \[ \pi^2(q)=-q, \; \pi^2(p)=-p - r_0 z, \; \pi^2(\alpha_i)=
\alpha_{i+2}, \; \pi^2(\epsilon_0)=-\epsilon_0
\]
as well as commutativity  rules:
\begin{equation}   s_i s_{i+2}=s_{i+2}  s_i, \; \pi_i \pi_{i+2}=\pi_{i+2}  \pi_i, 
\;  \pi_i s_{i+2}=\pi_{i+2}  s_i,, \; i=0,2 \, .
\lab{prop2}
\end{equation}
Relations \rf{prop1} and \rf{prop2}  amount to the 
following Coxeter group relations :
\begin{xalignat}{2}
\left( s_0 s_2 \right)^2 &=1,\; &\left( \pi_0 \pi_2 \right)^2 =1\,,
\lab{Back1}\\
\left( \pi_0 s_2 \right)^2 &=1,\; &\left( \pi_2 s_0 \right)^2 =1\, .
\lab{Back2}
\end{xalignat}
In the setting of P$_{III}$ model it is possible to realize $W [s_0,s_2,\pi_0,\pi_2,\pi^2]$ 
symmetry as an extended affine Weyl group  $C^{(1)}_2$ \cite{AGZ2016}.

Ince's equation  $I_{12}$ :
\begin{equation}
I_{12} \; :\,y_{xx}=  \frac{y_x^2}{y}+\mathcal{A} y^3+ \mathcal{B} y^2
+\mathcal{C}+\frac{\mathcal{D} }{y} 
\lab{Ince:12}
\end{equation}
as well as the incomplete $I_{38}$ (with $\mathcal{D}=0$) and the incomplete 
$I_{49}$ (with $\mathcal{A}=1$) are 
obtained from  P$_{III-V}$ for either either of two values of 
parameters given in \rf{special-limits}. 
For these models the symmetry
is still given by  $W [s_0,s_2,\pi_0,\pi_2,\pi^2]$ (or 
$W [s_1,s_3,\pi_1,\pi_3,\pi^2]$). However for parameters 
in \rf{special-limits}
actions of $\pi_i$ on $\alpha_j$ become identical to those of
$s_i$, and connection with the affine extended Weyl group $C^{(1)}_2$ 
realization can no longer be established
\cite{AAGZ2}.


We will illustrate the above comments by providing a brief derivation
of relevant Ince's equations for a first choice 
$r_1=0, J=1$ of the parameters among two listed in \rf{special-limits}.
Defining
\[ x =\ln z, \qquad w= q z, \qquad f = p/z
\]
we obtain from \rf{qzpz} equations:
\begin{equation}
\begin{split}
 w_x &= w^2 \left(2f+r_0 \right)- 
\left(\alpha_1+ \alpha_3-1 \right) w
+\epsilon_0 r_0 \, , \\
f_x &=   f \left(f+ r_0 \right)
\left( - 2 w \right) + 
(\alpha_1+\alpha_3-1) f - 
\alpha_2  r_0 \, ,
\end{split}
\lab{wfz}
\end{equation}
that also can be reproduced as Hamilton equations following from a
Hamiltonian :
\begin{equation}
H_{12}= f^2w^2+w^2fr_0+(\alpha_1+\alpha_3-1) f w+ \epsilon_0 r_0 f+\alpha_2
r_0 w \, .
\lab{Hfw}
\end{equation}
Eliminating $f$ one obtains the second order equation for $w$ :
\[
w_{xx}= \frac{w_x^2}{w}+w^3r_0^2+w^2r_0(\alpha_2-\alpha_0)
-\epsilon_0 r_0 (\alpha_0+\alpha_2)-\frac{\epsilon_0^2 r_0^2 }{w} \, ,
\]
in which we recognize equation I$_{12}$  of Ince \rf{Ince:12}.
The second order equation for $f$ written in terms of $y$ such that
\[ f= -\frac{r_0 y}{y-1}
\]
leads to Ince's equation I$_{38}$  \rf{Ince:38} 
with $\mathcal{A}=\alpha_0^2/2$, $\mathcal{B}
=-\alpha_2^2/{2}$, $\mathcal{C}=-2 \epsilon_0 r_0^2$
and  $\mathcal{D}=0$ and thus the equation obtained in this limit is
only an incomplete version of I$_{38}$  equation  \rf{Ince:38}. 

We now turn our attention to the remaining 
case listed in \rf{special-limits}:  $r_0 = 0$ and $J=-1$.
Inserting these values directly in \rf{qzpz} with $r_0=0$  yields 
(for $x =\ln z$)
\[
\begin{split}
q_x&= q \left(q-r_1 \right) 2p- 
\left(\alpha_1+\alpha_3 \right) q
+\alpha_1 r_1  \\
 p_x &=   p^2
\left(r_1 - 2q \right) + 
(\alpha_1+\alpha_3) p - 
\epsilon_1 r_1 
\end{split}
\]
Let us set $q=w,p=f$ and note that the Hamiltonian that
reproduces the above equations is:
\begin{equation}
{\bar H}_{12} = f^2w^2-wf^2 r_1-(\alpha_1+\alpha_3) f w+ \epsilon_1 r_1 w+\alpha_1
r_1 f\, ,
\lab{HfwC0}
\end{equation}
(note that the major difference from $H_{12}$ in \rf{Hfw} is the 
term $w f^2$ instead for $w^2f$.)

For the quantity $f=p$ we find from the above equations 
a second order equation: 
\[
f_{xx}= \frac{f_x^2}{f}+f^3 r_1^2+f^2r_1(-\alpha_1+\alpha_3)
+\epsilon_1 r_1 (\alpha_1+\alpha_3)-\frac{\epsilon_1^2 r_1^2 }{f}\, ,
\]
in which we again recognize the XII-th equation of Ince \rf{Ince:12}.
Furthermore we derive:
\[
\begin{split}
w_{xx}&= \frac{w_x^2}{2}\left( \frac{1}{w}+\frac{1}{w-r_1}\right)-2
r_1 \epsilon_1 w^2 +\alpha_1^2 r_1\\
&+\frac{2 r_1^2 \epsilon_1 w^2}{w-r_1}-
w r_1 \frac{\alpha_1^2+\alpha_3^2+4 r_1^2 \epsilon_1}{2(w-r_1)}
+ \frac{r_1^3 \alpha_1^2}{2(w-r_1)} \, .
\end{split}
\]
Defining $y$ in terms of $w$ as
\[
y= \frac{w}{w-r_1} \quad \text{or}  \quad w = \frac{r_1y}{y-1} \, ,
\]
we obtain
a special incomplete case of Ince's equation XLIX \rf{Ince:49} 
with the parameters $\mathcal{A}=1,
\mathcal{B}=\alpha_3^2/2,\mathcal{C}=-\alpha_1^2/2$
and $\mathcal{D}+\mathcal{E}=2 r_1^2 \epsilon_1$.

\section{Completing Hamiltonian \rf{Hfw} to obtain Ince's  eq. 38}
To obtain a Hamiltonian that would reproduce a 
complete equation I$_{38}$ 
we symmetrize Hamiltonian structures $H_{12}$ from \rf{Hfw} and ${\bar H}_{12}$
from \rf{HfwC0} by adding a 
term $ w f^2$ to the Hamiltonian $H_{12}$ :
\begin{equation}
H_{38} = f^2w^2+ \alpha w^2f +\kappa w f^2 +\beta f w+ \gamma f+\delta
w\, ,
\lab{hamkappa}
\end{equation}
where we also replaced $r_0$ by $\alpha$, $-(\alpha_1+\alpha_3-1)$ by
$\beta$, $\epsilon_0 r_0 $ by $ \gamma$ and $\alpha_2
r_0$ by $\delta$ to have a more general expression.
The  corresponding Hamilton equations are
\begin{equation}
\begin{split}
 w_x &= 2 w^2 f +\alpha w^2+ \beta  w + 2 \kappa w f
 + \gamma\, , \\
f_x &=   -2 w f^2 -2 \alpha wf -\beta f 
-\delta -\kappa f^2 \, .
\end{split}
\lab{kappawfz}
\end{equation}
Taking a second derivative of the first equation in \rf{kappawfz} and
eliminating $f$ one obtains a second-order differential equation 
for $w$ which agrees with I$_{38}$ in \rf{Ince:38} with the parameters:
\[ 
\begin{split}
{\cal A} &= \frac{1}{2 \kappa^2} (\alpha^2\kappa^4+\beta^2\kappa^2 -2 \alpha
\kappa^3 \beta+2 \alpha \kappa^2 \gamma +\gamma^2-2\beta \kappa \gamma) 
,\;\;{\cal B}= -\frac{\gamma^2}{2 \kappa^2} \\
{\cal C}&= \frac{\kappa}{2} (-4 \delta+2 \alpha \beta-3 \alpha^2 \kappa),
\qquad 
{\cal D}= - \kappa^2 \alpha^2 \, ,
\end{split}
\]
when using the variable $y$ : 
\[ y=\frac{w}{w+\kappa}\,.
\]
Similarly obtaining a second-order differential equation for $f$ from  \rf{kappawfz}
and defining $y$ through $f=-\alpha y/(y-1)$ will also yield equation
\rf{Ince:38}. Due to the fact that addition of 
$ \kappa w f^2$ term rendered the system  \rf{hamkappa} explicitly  symmetric 
in $w,f$ the  Hamilton  equations \rf{kappawfz} are invariant
 under $w,f$ rotation  here referred to as an $R$ operation :
 \begin{equation}
 \begin{split}
 R(w)&=-f,\; R(f)=-w, \; R(\kappa)=-\alpha, \; R(\alpha)=-\kappa\\
 R(\gamma)&=-\delta  \;R(\delta)=-\gamma , R (\beta)=\beta,
 R(x)=-x\, .
 \lab{Roperation}
\end{split}
\end{equation}
Obviously $R^2=1$.  Also the Hamilton equations \rf{kappawfz} are invariant under
 \begin{equation}
 \begin{split}
  \pi^2(w)&=-w, \; \pi^2(f)=-f - \alpha,   \; \; \pi^2(\alpha)=
 \alpha, \; \;  \pi^2(\beta)=\beta - 2 \kappa \alpha,\\
 \pi^2(\gamma)&= -\gamma, \; \; 
 \pi^2(\kappa)=-\kappa, \;  \; 
\pi^2(\delta)= - \delta +\alpha \beta -\kappa \alpha^2\, ,
\lab{pisquare}
\end{split}
\end{equation}
which also squares to $1$.
In addition the model is also invariant under modified transformations \rf{s0def} and
 \rf{s2def}:
 \begin{equation}
 \begin{split}
 s_2(w)&= w+ \frac{\delta}{\alpha f}, \quad  
 s_2 (f)= f, \quad
 s_2(\alpha) =  \alpha, \;\; s_2 (\kappa)=\kappa\\ 
 s_2 (\beta)&=\beta- 2 \frac{\delta}{\alpha}  , \;\;
 s_2(\gamma)=  \gamma -\frac{\kappa \delta}{\alpha} , \;\,
 s_2 (\delta)= -\delta
 \end{split}
 \lab{s2defk}
 \end{equation}
 and
 \begin{equation}\begin{split}
 s_0(w)&= w+ \frac{\beta-\delta/\alpha-\kappa \alpha}{f+\alpha},\: \; 
 s_0 (f)= f, \;\; s_0(\alpha)= \alpha, \;\; s_0(\kappa)=\kappa\\
 s_0(\beta) &=-\beta+2\frac{\delta}{\alpha}+2 \kappa \alpha,  \;\;
 s_0 (\gamma)=\gamma -\kappa ( \beta-\frac{\delta}{\alpha}-\kappa \alpha), 
\; \;
 s_0(\delta)=  \delta 
\end{split}
 \lab{s0defk}
 \end{equation}
 connected to each other via $s_0=\pi^2 s_2 \pi^2$ 
 and both being a symmetries of the 
 Hamilton equations \rf{kappawfz}.

 We can also define transformations
 \[ 
 \mathcal{S}_2=Rs_2R, \;\mathcal{S}_0=Rs_0R,
 \]
 with an explicit action for $\mathcal{S}_0$ and $\mathcal{S}_2$ being :
 \begin{equation}\begin{split}
 \mathcal{S}_2(w)&= w, \; \mathcal{S}_2 (f)= f+ \frac{\gamma}{w \kappa
 }, \;  \mathcal{S}_2(\beta) =\beta- 2 \gamma/ \kappa ,  \\
 \mathcal{S}_2(\delta)&=  \delta -\gamma \alpha/ \kappa, \,
 \mathcal{S}_2 (\gamma)=-\gamma, \; \mathcal{S}_2 (\alpha)=\alpha, 
 \end{split}
 \lab{S22def}
 \end{equation}
  \begin{equation}\begin{split}
 \mathcal{S}_0(w)&= w, \; \mathcal{S}_0 (f)= f+
 \frac{(\beta -\frac{\gamma}{\kappa}-\kappa \alpha)}{w+\kappa}
 \\
 \mathcal{S}_0(\beta) &=-\beta+2
 \gamma/ \kappa +2 \kappa \alpha , \; \mathcal{S}_0(\kappa)=\kappa \\
 \mathcal{S}_0(\delta)&=  \delta -\alpha(\beta-\frac{\gamma}{\kappa} -\kappa\alpha) ,
 \;
 \mathcal{S}_0 (\gamma)=\gamma  , \; \mathcal{S}_0(\alpha)
=\alpha \, ,
\end{split}
 \lab{SS00def}
 \end{equation}
with both transformations keeping the Hamilton equations \rf{kappawfz} invariant.
We note that $w$ remains invariant under actions of 
$\mathcal{S}_0, \mathcal{S}_2$ while  $f$ remains invariant under actions of
$s_0,s_2$.

In conclusion addition of an additional cubic term to Hamilton structures 
\rf{Hfw} or \rf{HfwC0} associated with I$_{12}$  
yields a Hamilton  structure \rf{hamkappa} of I$_{38}$  with a symmetry group no
longer involving the automorphisms $ \pi_i, i=0,2$. The manifest symmetry
between $f$ and $w$ variables gives raise to a new $w-f$ rotation $R$.

\section{Hamiltonian for Ince's I$_{49}$ and its symmetries}

We will here derive Ince's equation I$_{49}$  \rf{Ince:49} from the 
Hamilton function :
\begin{equation}
H_{49} = kf^3w^2+f^2 w^2 +\alpha f w^2 +\kappa w f^2
+\beta f w+ \gamma f +\delta w\, ,
\lab{ham49}
\end{equation}
where we allowed for the first time a term of the 5-th power in 
$f,w$ : $k f^3 w^2$ in addition to terms already present in $H_{38}$ 
\rf{hamkappa}.
A term of this dimension appears in the Hamiltonian
of the Painlev\'e VI equation.  
The corresponding Hamilton equations are 
\begin{equation}
\begin{split}
w_x&=+3kf^2w^2+w^2(2f+\alpha )+2\kappa wf+\beta w+
\gamma \, ,\\
f_x&=-2kf^3w+f(f+\alpha)(-2w)-\kappa f^2 -\beta f-\delta \, , 
\end{split}
\lab{ham49eqs}
\end{equation}
leading to Ince's equation I$_{49}$  \rf{Ince:49} when
coefficients $k$ and $\alpha$ are related through the condition:
\begin{equation}
k=-(\alpha+1), \quad \alpha \ne -1
\lab{conditionr0k}
\end{equation}
or $\alpha=-(k+1)$. The  coefficients $\mathcal{A},\mathcal{B},
\mathcal{C}, \mathcal{E}, \mathcal{D}$ of
equation \rf{Ince:49} are given in terms of coefficients $\kappa,
\alpha, \beta, \gamma, \delta$ from
\rf{ham49} as follows :
\begin{equation}
\begin{split}
\mathcal{A}&= -\frac{\alpha}{\alpha+1},\quad
\mathcal{C}= \frac12  \delta^2 (\alpha+1)/\alpha,\quad
\mathcal{B}=2 \gamma (\alpha+1) + \kappa^2/2\\
\mathcal{D} &= -\frac{\alpha+1}{2(2\alpha+1)} \left(\delta
+\beta+\kappa\right)^2,\;\,
\mathcal{E} =-
\frac{\left((1+\alpha ) (\delta +\alpha  (-\beta +\delta ))+
\alpha ^2 \kappa \right)^2}{2 \alpha  (1+\alpha )^2 (1+2 \alpha )} \, .
\end{split}
\lab{ham49coeffs}
\end{equation}

The parameter $\alpha$ needs to have values different from 
$=-1,-1/2,0,$ to avoid that 
$\mathcal{A}=0$ or $\mathcal{A}=1$ or $\mathcal{D}$ or $\mathcal{E}$ 
becoming infinite.

As we will see below
the condition \rf{conditionr0k} required so that the Hamilton equations
derived  from \rf{ham49} would reproduce equation I$_{49}$  in
\rf{Ince:49} also enables  several symmetries
of \rf{ham49} system. In the discussion below we will assume that the
condition \rf{conditionr0k}
holds and consider those symmetries that maintain this relation.

One symmetry transformation that keeps equations \rf{ham49eqs}
invariant is
\begin{equation}
 \begin{split}
 s_2(w)&= w+ \frac{\delta}{\alpha f}, \quad  
 s_2 (f)= f\\
 s_2(\alpha) &=  \alpha, \; s_2 (\delta)= -\delta,\; \; s_2 (\beta)=\beta- 2 \frac{\delta}{\alpha}  \\
 s_2(\gamma)&=  \gamma -\frac{\kappa \delta}{\alpha}-(\alpha+1)
 \frac{\delta^2}{\alpha^2} , \,
 \;\; s_2 (\kappa)=\kappa +2 (\alpha+1) \frac{\delta}{\alpha} \, ,
 \end{split}
 \lab{s2def49}
 \end{equation}
with
\[s_2^2=1 \,.
\]

For equations \rf{ham49eqs} with condition $k=-\alpha-1$ satisfied
we can also define additional symmetry $\pi_0$ made possible by
presence of term $k f^3 w^2$ in addition to $\alpha f w^2$ in $H_{49}$. 
This symmetry is defined 
as follows
\begin{equation}\begin{split}
\pi_0(f)&= -\frac{\alpha}{(\alpha+1) f}, \quad  
\pi_0 (w)=  \left( \frac{\alpha+1}{\alpha} f^2w+ \frac{X}{\alpha} f\right)\\
\pi_0 (\beta) &=  -\beta+2 \frac{X}{\alpha+1},\; \pi_0
(\kappa)= \frac{1}{\alpha} (-2 X \alpha + \delta (\alpha+1))\\
\pi_0 (\delta)&=  \frac{\alpha}{\alpha+1} (\kappa +2 X), \;
\pi_0 (k)=k, \; \pi_0 (\alpha)=\alpha\\
\pi_0(\gamma)&=-\gamma +\frac{X(\delta(\alpha+1)+\kappa
\alpha)}{\alpha(\alpha+1)} \, ,
\end{split}
\lab{pi01k}
\end{equation}
in terms of $X$ being  a root of a quadratic 
equation:
\begin{equation}
-\gamma (\alpha+1)+\kappa X+X^2=0\, .
\lab{rootX}
\end{equation}
Acting with $\pi_0$ directly
on $X=(-\kappa\pm \sqrt{\kappa^2+4 \gamma(\alpha+1)})/2$ 
yields
\[
\pi_0 (X)= X-\delta \frac{\alpha+1}{2\alpha} \pm \frac{1}{2}
\sqrt{\delta^2 (\alpha+1)^2/\alpha^2}\,.
\]
Thus there are two possible values for $\pi_0(X) $ :
\begin{equation}
\pi_0(X) = \left\{ \begin{matrix} X= \pi_{0\, +} (X) \\
X-\delta-\delta/\alpha = \pi_{0\,-} (X)\, ,
\end{matrix}\right.
\lab{pi0AB}
\end{equation}
where we associated two different transformations $\pi_{0\, +}$
and $\pi_{0\,-}$ to two possible actions of $\pi_0$ on $X$. Both 
$\pi_{0\, +}$ and $\pi_{0\,-}$ act on other quantities in accordance
with \rf{pi01k}.

Both transformations $\pi_{0\, +}$ and $\pi_{0\,-}$  keep  
equations \rf{ham49eqs} invariant and preserve the condition
$k=-(\alpha+1)$. Furthermore it holds  that :
\begin{equation} 
\pi_{0\, +}^2=1, \quad \pi_{0\,-}^4 =1 \, ,
\lab{pi0Asquare}
\end{equation}
with few intermediate explicit formulas being:
\begin{equation}
\begin{split}
\pi_{0\,-}^2 (w)&=s_2 (w),\; \pi_{0\,-}^2 (f)=s_2 (f), \;
\pi_{0\,-}^2 (\alpha)=s_2 (\alpha),\;\pi_{0\,-}^2 (\beta)=s_2
(\beta) 
\\
\pi_{0\,-}^2 (\delta)&=s_2 (\delta) ,  \; 
\pi_{0\,-}^2 (\kappa) = s_2 (\kappa) , \;
\pi_{0\,-}^2 (\gamma) = s_2 (\gamma)\\
\pi_{0\,-}^2 (X)&= -X- \frac{\delta(\alpha+1)}{\alpha}-\kappa\, .
\end{split}
\lab{pi0Bsquare}
\end{equation}
From relations \rf{pi0Bsquare}  and 
\begin{equation}
s_2 (X) = X -\delta \frac{\alpha+1}{\alpha} \, ,
\lab{s2X}
\end{equation} and $s_2 \pi_{0\,-}(X)=X$ it follows
that
\begin{equation}
(s_2 \pi_{0\, \pm})^4 =1\, .
\lab{s2pi0B4}
\end{equation}
Thus both $\pi_{0\, +}$ and $\pi_{0\,-}$ satisfy the relation
$(s_2 \pi_{0})^4 =1$ although only $\pi_{0\, +}$ squares to one! 
In addition $\pi_{0\, +}$ and $\pi_{0\,-}$ transformations 
satisfy the relation
\begin{equation}
(\pi_{0\,-}\pi_{0\, +})^2 =1,
\lab{p0pp0m}
\end{equation}
that can be rewritten equivalently as 
\begin{equation}
\pi_{0\,-}\pi_{0\, +} \pi_{0\,-} = \pi_{0\, +}, \qquad
\pi_{0\,-}^2\pi_{0\, +} \pi_{0\,-}^2 = \pi_{0\, +}, 
\lab{p0mp0pp0m}
\end{equation}
using relation \rf{pi0Asquare}.
The last identity can also be written
as 
\begin{equation}
(\pi_{0\,+}\pi_{0\, -}^2)^2 =1\,.
\lab{p0pp0mb}
\end{equation}

Equations \rf{ham49eqs} with the condition
$k=-(\alpha+1)$ are also invariant under transformations
of $\pi^2$ defined as
\begin{equation}
\begin{split}
\pi^2 (w) &= (2+3\alpha) w , \qquad  \;\; \pi^2 (f) = - f +1 \\
\pi^2 (k)&=-\frac{\alpha+1}{(2+3\alpha)},\; \;\; 
\pi^2 (\alpha)=-\frac{2 \alpha+1}{2+3\alpha}, \;\; 
\pi^2 (\beta)= \beta+2\kappa\\
\pi^2 (\kappa)&=-\kappa, \;\; \pi^2 ( \delta)=- \delta-\kappa-\beta
, \;\; 
\pi^2 (\gamma)=(2+3\alpha)\gamma
\end{split}
\lab{trans2rho}
\end{equation}
{}From
\[
 (2+3\alpha)  (2+3\pi^2(\alpha))=   (2+3\alpha)  (2-3
 \frac{2\alpha+1}{2+3\alpha})=1
 \]
it follows that \[(\pi^2)^2=1\] and since $\pi^2(k)=-\pi^2(\alpha)-1$  then
$\pi^2$   defined in \rf{trans2rho} is an automorphism that 
squares to $1$ and leaves \rf{ham49eqs} with the condition
$k=-(\alpha+1)$ invariant !

Define now 
\[
s_0 = \pi^2 s_2 \pi^2 \,,
\]
which obviously squares to one.
Then
\begin{equation}
 \begin{split}
s_0(w)&= w- \frac{\beta+\delta+\kappa}{(1+2\alpha) (f-1)}, \quad  
s_0 (f)= f\\
s_0(\alpha) &=  \alpha,\;  \; s_0 (\beta)=\frac{\beta- 2 \alpha
 (\delta+\kappa)}{1+2 \alpha}  \\
s_0(\gamma)&=\,{\frac {\gamma- \left( 1+\alpha \right)  
\left( -4\,\alpha\,\gamma+ \left( \delta+\beta \right)^{2} \right) 
- \left( \delta+\beta \right) \kappa
+\alpha\,{\kappa}^{2}}{ \left( 1+2\,\alpha \right)^{2}}}  , \\
s_0 (\delta)&= \delta, \;\; 
s_0 (\kappa)=-\frac{2(1+\alpha) (\beta+\delta)+\kappa}{1+2 \alpha}\,. 
 \end{split}
 \lab{s0def49}
 \end{equation}

It follows then that
\[
(s_2 s_0)^2 =1\, .
\]
For the transformation  
\[ \pi_{2\, \pm} = \pi^2 \pi_{0\, \pm} \pi^2\, ,
\]
we are able to find based on \rf{pi01k} and \rf{trans2rho}
the transformations rules :
\begin{equation}\begin{split}
\pi_2(f)&= \frac{\alpha+(\alpha+1)f}{(\alpha+1) (f-1)}, \quad  
\pi_2 (w)=  -\frac{1}{2\alpha+1} \left( (\alpha+1) (f-1)^2w+ 
\frac{X}{\alpha} (f-1)\right)\\
\pi_2 (\beta) &=  \frac{1}{2\alpha+1} \left(\beta-2 \alpha \kappa+2
\delta(\alpha+1) \right) -\frac{2X \alpha}{\alpha+1} 
\; ,\; \pi_2
(\kappa)= -(\delta+\kappa+\beta)\frac{\alpha+1}{2\alpha+1}-2 X 
\\
\pi_2 (\delta)&= -\delta + \frac{\alpha}{2\alpha+1} (\delta+\beta)
-\kappa \frac{\alpha^2}{(2 \alpha+1)(\alpha+1)}, \;
 \; \pi_2 (\alpha)=\alpha\\
\pi_2(\gamma)&=-\gamma +X \left( \frac{\delta+\beta}{2\alpha+1}
+\kappa ( \frac{1}{2\alpha+1}-\frac{1}{\alpha+1})\right) \,.
\end{split}
\lab{pi21k}
\end{equation}
In addition it holds
\begin{equation}
\pi_2(X) = \left\{ \begin{matrix} X= \pi_{2\, +} (X) \\
X+(\delta+\kappa+\beta)\frac{\alpha+1}{2\alpha+1} = \pi_{2\,-} (X) \, .
\end{matrix}\right.
\lab{pi2AB}
\end{equation}
Furthermore it follows that 
\[
\pi_{2\,+}^2=1,\qquad \pi_{2\,-}^4=1
\]
and also
\[
(s_0 \pi_{2\, \pm})^4=1\, .
\]
We also find from relations \rf{p0pp0m} and \rf{p0pp0mb}
that:
\begin{equation}
(\pi_{2\,-}\pi_{2\, +})^2 =1, \qquad
(\pi_{2\,+}\pi_{2\, -}^2)^2 =1\,.
\lab{p2pp2m}
\end{equation}
In conclusion adding a $5$-th power term to Hamilton structure
 \rf{hamkappa} of I$_{38}$  
yielded Hamilton  structure \rf{ham49} of I$_{49}$ and restored $\pi_0$
symmetry that was absent in $H_{38}$.
Remarkably,  the underlying symmetry group contains transformations
$s_0, s_2, \pi_0,
\pi_2$ and $\pi^2$ of $W[ s_0,s_2,\pi_0, \pi_2, \pi^2]$. 
However the Coxeter group relations are in some cases (e.g \rf{s2pi0B4} 
and \rf{p2pp2m}) of higher order as compared with 
simple Coxeter group relations \rf{Back1} and \rf{Back2} of
symmetry transformations of P$_{III}$ or I$_{12}$ models.

\section{Conclusions}
We presented here a study of Hamiltonian structures of I$_{12}$,
I$_{38}$ and I$_{49}$ and their symmetries. 
The mixed  P$_{III-V}$ equations taken for various
special values of the underlying parameters provided a useful starting
point for this analysis.

The Hamiltonian structure of I$_{12}$ shared its symmetry generators and 
underlying Coxeter relations with P$_{III}$ equation 
although its symmetry generators acted differently on a set parameters 
hindering its interpretation as an extended affine Weyl group as
described in details in \cite{AAGZ2}.
To obtain I$_{38}$ and I$_{49}$ hamiltonian structures extra terms needed to be 
added to the Hamiltonian for  I$_{12}$. 
For I$_{38}$ that resulted in additional symmetry rotation operation
and a totally different content of the underlying symmetry group.
For I$_{49}$ model the underlying symmetry structure 
contains the same generators as those of P$_{III}$ 
or  I$_{12}$ but the added higher dimensional term in the Hamiltonian
resulted in different higher order Coxeter relations among symmetry
generators. 

\ack
JFG and AHZ thank CNPq and FAPESP for financial support. VCCA thanks grant 2016/22122-9, S\~ao Paulo Research Foundation (FAPESP) for financial support.

\end{document}